\documentclass[reprint,amssymb,amsmath,superscriptaddress,aps,showpacs,10pt,floatfix,longbibliography,prx]{revtex4-1}

\usepackage[pdftex]{graphicx} 
\usepackage{epstopdf}
\usepackage{verbatim}
\usepackage{amsmath}
\usepackage{color}
\usepackage{subfigure}
\usepackage{amsbsy}
\usepackage{wasysym}

\begin{document}

\title{Evidence for a gapless Dirac spin-liquid ground state\\in a spin-3/2 triangular-lattice antiferromagnet}

\author{Jiabin Liu}
\thanks{These authors contributed equally to this work}
\affiliation{Wuhan National High Magnetic Field Center and School of Physics, Huazhong University of Science and Technology, 430074 Wuhan, China}

\author{Benqiong Liu}
\thanks{These authors contributed equally to this work}
\affiliation{Key Laboratory of Neutron Physics, Institute of Nuclear Physics and Chemistry, CAEP, Mianyang 621900, China}

\author{Long Yuan}
\thanks{These authors contributed equally to this work}
\affiliation{Wuhan National High Magnetic Field Center and School of Physics, Huazhong University of Science and Technology, 430074 Wuhan, China}

\author{Boqiang Li}
\affiliation{Wuhan National High Magnetic Field Center and School of Physics, Huazhong University of Science and Technology, 430074 Wuhan, China}

\author{Lei Xie}
\affiliation{Key Laboratory of Neutron Physics, Institute of Nuclear Physics and Chemistry, CAEP, Mianyang 621900, China}

\author{Xiping Chen}
\affiliation{Key Laboratory of Neutron Physics, Institute of Nuclear Physics and Chemistry, CAEP, Mianyang 621900, China}

\author{Hongxia Zhang}
\affiliation{Department of Physics and Beijing Key Laboratory of Opto-electronic Functional Materials and Micro-nano Devices, Renmin University of China, Beijing 100872, P. R. China}

\author{Daye Xu}
\affiliation{Department of Physics and Beijing Key Laboratory of Opto-electronic Functional Materials and Micro-nano Devices, Renmin University of China, Beijing 100872, P. R. China}

\author{Wei Tong}
\email{weitong@hmfl.ac.cn}
\affiliation{Anhui Province Key Laboratory of Condensed Matter Physics at Extreme Conditions, High Magnetic Field Laboratory, Chinese Academy of Sciences, Hefei 230031, China}

\author{Jinchen Wang}
\email{jcwang\_phys@ruc.edu.cn}
\affiliation{Department of Physics and Beijing Key Laboratory of Opto-electronic Functional Materials and Micro-nano Devices, Renmin University of China, Beijing 100872, P. R. China}

\author{Yuesheng Li}
\email{yuesheng\_li@hust.edu.cn}
\affiliation{Wuhan National High Magnetic Field Center and School of Physics, Huazhong University of Science and Technology, 430074 Wuhan, China}

\date{\today}

\begin{abstract}
We report a comprehensive investigation of the magnetism of the $S$  = 3/2 triangular-lattice antiferromagnet, $\alpha$-CrOOH(D) (delafossites green-grey powder). The nearly Heisenberg antiferromagnetic Hamiltonian ($J_1$ $\sim$ 23.5 K) with a weak single-ion anisotropy of $|D|$/$J_1$ $\sim$ 4.6\% is quantitatively determined by fitting to the electron spin resonance (ESR) linewidth and susceptibility measured at high temperatures. The weak single-ion anisotropy interactions, possibly along with other perturbations, e.g. next-nearest-neighbor interactions, suppress the long-range magnetic order and render the system disordered, as evidenced by both the absence of any clear magnetic reflections in neutron diffraction and the presence of the dominant paramagnetic ESR signal down to 2 K ($\sim$ 0.04$J_1$$S^2$), where the magnetic entropy is almost zero. The power-law behavior of specific heat ($C_m$ $\sim$ $T^{2.2}$) observed below the freezing temperature of $T_f$ = 25 K in $\alpha$-CrOOH or below $T_f$ = 22 K in $\alpha$-CrOOD is insensitive to the external magnetic field, and thus is consistent with the theoretical prediction of a gapless \emph{U}(1) Dirac quantum spin liquid (QSL) ground state. At low temperatures, the spectral weight of the low-energy continuous spin excitations accumulates at the K points of the Brillouin zone, e.g. $|\mathbf{Q}|$ = 4$\pi$/(3$a$), and the putative Dirac cones are clearly visible. Our work is a first step towards the understanding of the possible Dirac QSL ground state in this triangular-lattice magnet with $S$ = 3/2.
\end{abstract}

\maketitle

\section{Introduction}

Periodic order of spins (electronic magnetic moments) is usually stabilized by the interactions in the periodic crystalline lattice of a magnet at low temperatures. When geometrical frustration kicks in, not all interactions can be satisfied simultaneously, the periodic spin order may get suppressed even at zero temperature, and exotic disordered phases, e.g. quantum spin liquid (QSL), may emerge~\cite{wannier1950antiferromagnetism,balents2010spin,RevModPhys.89.025003}. The prototype of a QSL, i.e. resonating-valence-bond state, had been proposed by Anderson on the $S$ = 1/2 triangular-lattice Heisenberg antiferromagnetic (THAF) model in 1973~\cite{ANDERSON1973153}. Since then, QSL has been arousing great interest in the community owing to its novel properties, such as quantum number fractionalization and intrinsic topological order, which may be intimately related to the understanding of high-temperature superconductivity and the realizing of topological quantum computation~\cite{anderson1987resonating,PhysRevLett.59.2095,RevModPhys.80.1083,broholm2020quantum}.

Although the QSL ground state is not realized in the ideal $S$ = 1/2 nearest-neighbor (NN) THAF model as expected~\cite{PhysRevLett.60.2531,PhysRevLett.68.1766,PhysRevLett.82.3899}, slight modifications of the ideal model, i.e. the inclusion of further-nearest-neighbor interactions~\cite{doi:10.7566/JPSJ.83.093707,PhysRevB.92.041105,PhysRevB.91.014426,PhysRevLett.123.207203}, multiple-spin exchanges~\cite{PhysRevB.60.1064,PhysRevB.72.045105}, coupling anisotropy~\cite{PhysRevLett.120.207203}, or (and) randomness~\cite{doi:10.7566/JPSJ.83.034714,PhysRevLett.119.157201,wu2020Exact}, etc., can give rise to various QSL ground states. On the experimental side, there exist lots of $S$ = 1/2 triangular-lattice antiferomagnets (TAFs) exhibiting QSL behaviors, such as $\kappa$-(BEDT-TTF)$_2$Cu$_2$(CN)$_3$~\cite{PhysRevLett.91.107001,yamashita2008thermodynamic,2008Thermal}, EtMe$_3$Sb[Pd(dmit)$_2$]$_2$~\cite{PhysRevB.77.104413,PhysRevB.84.094405,2010Instability,Yamashita11Gapless}, Ba$_3$CuSb$_2$O$_9$~\cite{PhysRevLett.106.147204}, and YbMgGaO$_4$~\cite{li2015gapless,li2015rare,PhysRevLett.117.097201,shen2016spinon,paddison2016continuous,PhysRevLett.118.107202,li2017nearest,PhysRevLett.122.137201,doi:10.1002/qute.201900089}. Furthermore, QSLs can also be stabilized in $S$ = 1 TAFs with some additional interaction terms~\cite{PhysRevB.81.224417,PhysRevB.84.180403,PhysRevLett.108.087204,PhysRevB.86.224409}, which may be well materialized in NiGa$_2$S$_4$~\cite{Nakatsuji2005Spin,Nakatsuji2010Novel} and Ba$_3$NiSb$_2$O$_9$~\cite{PhysRevLett.107.197204}. Like the $S$ = 1/2 TAF, the $S$ = 3/2 TAF also has odd number of spin-half electrons per unit cell, and thus its ground state must be unconventional as long as the time reversal symmetry is preserved, i.e. the spin system keeps disordered, according to the Hastings-Oshikawa-Lieb-Schultz-Mattis theorem~\cite{Watanabe2015Filling,PhysRevB.69.104431,PhysRevLett.84.1535,Lieb1966Two}. However, the evidence of QSLs in a perfect $S$ = 3/2 TAF compound without spin dilution have been rarely reported so far, despite the recent theoretical proposal of a Kiteav QSL on the $S$ = 3/2 honeycomb lattice~\cite{PhysRevLett.124.087205}

In this paper, we perform a thorough investigation of the frustrated magnetism of the $S$ = 3/2 TAF compound $\alpha$-CrOOH(D), including specific heat, magnetization (susceptibility), X-band \& high-field electron spin resonance (ESR), elastic \& inelastic neutron scattering measurements, as well as the finite-temperature Lanczos diagonalization (FLD) simulation. $\alpha$-CrOOH(D) has a well separated and regular triangular lattice of Cr$^{3+}$ ($S$ = 3/2) ions with symmetrically forbidden Dzyaloshinsky-Moriya (DM) interactions and with negligible magnetic impurities. Through a combined fit to the high-temperature ESR linewidth and susceptibility, the nearly THAF Hamiltonian with a weak single-ion anisotropy is quantitatively determined. Both the presence of the main paramagnetic ESR signal and the absence of any clear magnetic peaks in neutron diffraction at low temperatures indicate the ground state of the spin system is nearly symmetric, which can be roughly understood by including of the weak single-ion anisotropy ($D$/$J_1$ $\sim$ -4.6\%) and (or) other potential perturbations, e.g. next-nearest-neighbor (NNN) Heisenberg interactions. Below $T_f$, the measured power-law temperature dependence of specific heat, $C_m$ $\sim$ $T^{2.2}$, is well consistent with the theoretical prediction of a gapless QSL with Dirac nodes. Furthermore, the low-energy magnetic continuum of $\alpha$-CrOOD accumulates at the K points of the Brillouin zone, e.g. $|\mathbf{Q}|$ = 4$\pi$/(3$a$), possibly confirming the formation of the gapless Dirac nodes at low temperatures. 

\section{Technical details}

Powder samples of $\alpha$-CrOOH and $\alpha$-CrOOD were synthesized using the high-temperature (near the critical temperature of water, 374 $^o$C) hydrothermal technique at an autogenic pressure of $\leq$ 20 MPa~\cite{A1976Hydrothermal,Ichikawa1999Powder}. A 50 mL stainless steel liner was charged with 15 mmol Na$_2$CrO$_4$$\cdot$4H$_2$O ($>$ 99\%, Adamas Reagent Co., Ltd), 22.5 mmol CHNaO$_2$ ($>$ 98.5\%, Shanghai Titan Scientific Co., Ltd), and 10 mL of high-purity H$_2$O (for $\alpha$-CrOOH) or D$_2$O ($\alpha$-CrOOD). The liner was capped and mounted into a stainless steel pressure vessel. The vessel was heated to 350 $^o$C at a rate of 1 $^o$C/min, and the temperature was maintained for 20 hours and then cooled down to room temperature at a rate of -0.5 $^o$C/min. A green-grey powder ($\sim$ 1.1 g, the yield of $\sim$ 85\%) of $\alpha$-CrOOH or $\alpha$-CrOOD was obtained, and the phase purity of sample was confirmed by both x-ray and neutron diffraction measurements (Appendix~\ref{a1}). The successful replacement of H by D was confirmed by the significant difference of the neutron diffraction spectra owing to the neutron scattering length difference of H and D [please compare Fig.~\ref{figs1} (b) to Fig.~\ref{fig3} (a)], and by the change of the freezing temperature from 25 K in $\alpha$-CrOOH to 22 K in $\alpha$-CrOOD [see Fig.~\ref{figs2} (b)]~\cite{Matsuo2006Isotope,maekawa2007deuteration-induced}.

X-ray powder diffraction measurements were performed using a diffractometer (MiniFlex600, Rigaku). Direct-current (dc) magnetization and susceptibility of $\alpha$-CrOOH and $\alpha$-CrOOD were measured down to 1.8 K using powders of $\sim$ 40 mg, by a magnetic property measurement system (MPMS, Quantum Design, up to 7 T) and by a physical property measurement system (PPMS, Quantum Design, up to 14 T). The specific heat (2 $\leq$ $T$ $\leq$ 200 K) was measured at 0, 2, and 4 T using a dry-pressed disk of powder ($\sim$ 4 mg) in a PPMS. N grease was used to facilitate thermal contact between the sample and the puck, and the sample coupling was better than 99\%. The specific-heat contributions of the grease and puck under different applied fields were measured at first and subtracted from the data. Both steady-high-field (high-frequency) and X-band ESR measurements ($\sim$ 30 mg of $\alpha$-CrOOH) were performed on the Steady High Magnetic Field Facilities, Chinese Academy of Sciences. The high-field ESR data down to 2 K were collected by sweeping the steady magnetic field from 0 up to 20 T at selected frequencies of $f$ = 214, 230, and 331.44 GHz. The X-band ($f$ $\sim$ 9.4 GHz) ESR measurements were carried out using a Bruker EMX plus 10/12 continuous wave spectrometer. The measured ESR signals/modes are fitted using the derivative Lorentzian function on a powder average,
\begin{multline}
\frac{dA_{bs}}{\mu_0dH}=B_{ck}+\frac{16A_0\omega}{3\pi}\{\frac{2\mu_0(H_{res}^{ab}-H)}{[4\mu_0^2(H_{res}^{ab}-H)^2+\omega^2]^2}\\
+\frac{\mu_0(H_{res}^{c}-H)}{[4\mu_0^2(H_{res}^{c}-H)^2+\omega^2]^2}\},
\label{Eq1}
\end{multline}
where $A_0$ and $B_{ck}$ are the integral intensity and background, $\omega$ is the full width at half maximum (FWHM), $\mu_0H_{res}^{ab}$ = $hf$/($\mu_Bg_{ab}$) and $\mu_0H_{res}^{c}$ = $hf$/($\mu_Bg_{c}$) are the resonance fields along the $ab$ plane and $c$ axis, respectively. The neutron diffraction experiments were conducted with the high-intensity multi-section neutron powder diffractometer, Fenghuang~\cite{XIE201931}, using powder samples of $\sim$ 2.2 g, at China Mianyang Research Reactor (CMRR). The incident neutron wavelength of 1.5925 {\AA} was determined by the monochromator, Ge (5 1 1). The inelastic neutron scattering (INS) experiment was performed using $\alpha$-CrOOD of $\sim$ 6 g on the BOYA multiplexing cold neutron spectrometer stationed at the China Advanced Research Reactor (CARR). Utilizing the CAMEA concept~\cite{doi:10.1063/1.4943208}, BOYA simultaneously probes five fixed final energies ($E_f$) of 3.0 (energy resolution of $\sigma$ = 0.11 meV), 3.5 ($\sigma$ = 0.15 meV), 4.0 ($\sigma$ = 0.18 meV), 4.5 ($\sigma$ = 0.21 meV) and 5.0 meV ($\sigma$ = 0.26 meV), and 17--34 scattering angles spanning 120$^o$ in the horizontal plane. Highly oriented pyrolytic graphite (HOPG) crystals were used as the monochromator as well as the analyzers. Cryogenic Be filters were applied after the sample to cut off energies higher than 5 meV.

\begin{figure*}[t]
\begin{center}
\includegraphics[width=18cm,angle=0]{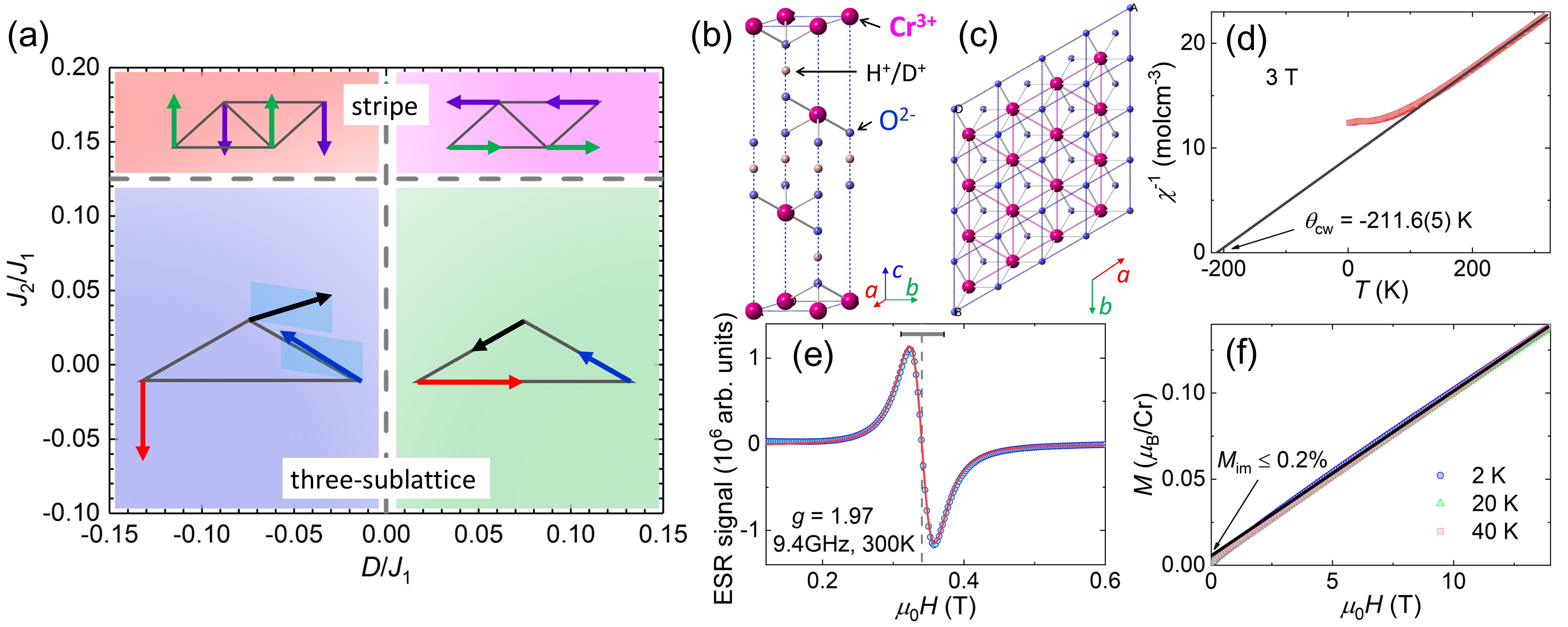}
\caption{(Color online)
(a) Classical phase diagram of the triangular-lattice antiferromagnet calculated by Monte Carlo simulation in the zero-temperature limit. (b) Crystal structure of $\alpha$-CrOOH(D). (c) Triangular lattice of Cr$^{3+}$ ions in the $ab$ plane, with showing NN O$^{2-}$ ions. The dashed blue lines in (b) and (c) mark the unit cells. (d) Inverse dc susceptibility measured at 3 T. The black line shows the linear fit to the data above 200 K. (e) X-band ($f$ = 9.4 GHz) ESR spectrum measured at 300 K. The red line shows the fit to the experimental data using the derivative Lorentzian function. The dashed line and horizontal bar present the fitted resonance field (Lorentzian center, $\mu_0H_{res}$) and linewidth (Lorentzian FWHM, $\omega$), respectively. (f) Magnetization measured at selected temperatures, with the black line showing the linear fit to the 2 K data above 3 T.}
\label{fig1}
\end{center}
\end{figure*}

We carried out the FLD calculations~\cite{Lanczos1950An,PhysRevB.49.5065} for the susceptibility~\footnote{We calculate the powder-averaged value, ($\chi_{\parallel}$+2$\chi_{\perp}$)/3, where $\chi_{\parallel}$ and $\chi_{\perp}$ are the susceptibilities parallel and perpendicular to the $c$ axis, respectively.}, specific heat, ESR linewidth, and static structure factor on the 12-site $S$ = 3/2 TAF clusters with periodic boundary conditions (PBC) (Appendix~\ref{a2})~\cite{PhysRevX.10.011007}. Then, we further performed the combined fit to the bulk susceptibility and ESR linewidth measured above 60 K by minimizing the residual,
\begin{equation}
R_p=\sqrt{\frac{1}{N}\sum_{j}(\frac{X_j^{obs}-X_j^{cal}}{\sigma_j^{obs}})^2},
\label{Eq2}
\end{equation}
where $N$, $X_j^{obs}$ and $\sigma_j^{obs}$ are the number of the data points, the measured value and its standard deviation, respectively, whereas $X_j^{cal}$ is the calculated value. The Monte Carlo simulation for the classical ground-state phase diagram was performed on the 144 (12$\times$12) -site cluster with PBC, by a gradual decrease of temperature from $J_1$ down to 0.001$J_1$ [see Fig.~\ref{fig1}(a)]. The powder INS spectra were calculated for the ideal $S$ = 3/2 THAF model by the Spinw-Matlab code based on the linear spin-wave theory~\cite{toth2015linear} [see Figs.~\ref{fig7}(d)-\ref{fig7}(f)]. An average of the results calculated along 10000 random orientations was performed at each $|\mathbf{Q}|$ for $S^{\perp}$($|\mathbf{Q}|$,$E$) with a high signal-to-noise ratio. The final INS intensities were calculated by a convolution of the instrumental resolution (Gaussian) function (FWHM of $\sigma$ = 0.25 meV) into the spectra, $I^{\perp}$($|\mathbf{Q}|$,$E$) = $|f(|\mathbf{Q}|)|^2\int$$S^{\perp}$($|\mathbf{Q}|$,$E'$)$\frac{\exp[-4\ln2(E-E')^2/\sigma^2]}{\sigma\sqrt{\pi/(4\ln2)}}$$dE'$, where $|f(|\mathbf{Q}|)|^2$ is the magnetic form factor of Cr$^{3+}$ in the dipole approximation. The international system of units is used throughout this paper.

\section{Quantum spin Hamiltonian}

Both $\alpha$-CrOOH and $\alpha$-CrOOD are good insulators with room-temperature resistance larger than 20 M$\Omega$. Through Rietveld refinements, both x-ray and neutron diffraction measurements indicate that the delafossites structure of the sample with the $R\bar{3}m$ space group maintains at least down to 8 K (Appendix~\ref{a1})~\footnote{Please note that $\alpha$-CrOOH (green-grey powder with the delafossites structure) investigated in this work is significantly different from HCrO$_2$ (dark brown powder with the ordered rock salt structure) reported by Hemmida \emph{et al.}~\cite{PhysRevB.80.054406}.}, as reported in Refs.~\cite{A1976Hydrothermal,Ichikawa1999Powder}. The crystal structure of $\alpha$-CrOOH(D) is shown in Fig.~\ref{fig1}(b), and the stacking pattern of O-H(D)-O layers is straight, which is similar to that of CuCrO$_2$ or AgCrO$_2$~\cite{PhysRevLett.101.067204}. The triangular layers of magnetic Cr$^{3+}$ ions are well spatially separated by nonmagnetic layers [Fig.~\ref{fig1}(b)] rendering the interlayer magnetic coupling weak. The high symmetry of the space group ensures the triangular lattice regular without spatial anisotropy [Fig.~\ref{fig1}(c)]. The inversion centers of the periodic crystalline lattice are located at halfway sites between neighboring Cr$^{3+}$ ions, and thus the antisymmetric DM interactions are forbidden by the $R\bar{3}m$ space group symmetry of $\alpha$-CrOOH(D)~\cite{PhysRevLett.4.228}.

Above $\sim$ 200 K, the Curie-Weiss behavior of bulk susceptibility is clearly observed in $\alpha$-CrOOH [Fig.~\ref{fig1}(d)]. The fitted Curie constant ($C$) indicates an effective magnetic moment, $g\mu_{B}\sqrt{S(S+1)}$ = $\sqrt{3k_BC/(N_A\mu_0)}$ = 3.861(2) $\mu_{B}$. Here, the $g$ factor is measured to be nearly constant and isotropic above 40 K by ESR, $g$ = 1.97 $\sim$ 2 [see Eq.~(\ref{Eq1}), Figs.~\ref{fig1}(e) and ~\ref{fig5}(c)], and thus $S$ $\sim$ 1.52 $\sim$ 3/2 is obtained. The 3$d$ electrons should be localized spatially owing to the poor conductivity, and the spin-orbit coupling of Cr$^{3+}$ is weak as evidenced by $g$ $\sim$ 2. Therefore, we can expect a nearly Heisenberg Hamitionian for $\alpha$-CrOOH(D) at 0 T as proposed in other Cr$^{3+}$-based TAFs of \emph{A}CrO$_2$ (\emph{A} = Li, Cu, and Ag)~\cite{collins1997review,Kadowaki1990Neutron},
\begin{multline}
\mathcal{H}=J_1\sum_{\langle jl\rangle}\mathbf{S}_j\cdot\mathbf{S}_l+\mathcal{H}'\\
=J_1\sum_{\langle jl\rangle}\mathbf{S}_j\cdot\mathbf{S}_l+D\sum_{j}(S_j^z)^2+J_2\sum_{\langle\langle jl'\rangle\rangle}\mathbf{S}_j\cdot\mathbf{S}_{l'},
\label{Eq3}
\end{multline}
where both single-ion anisotropy ($D$) and NNN interaction ($J_2$) terms are much weaker than the NN Heisenberg coupling ($J_1$). The fitted Weiss temperature [see Fig.~\ref{fig1}(d)], $\theta_{cw}$ = -211.6(5) K, roughly measures the dominant antiferromagnetic (AF) coupling by, $J_1$+$J_2$ $\sim$ -3$\theta_{cw}$/[6$S$($S$+1)] $\sim$ 28 K. In the classical level, the inclusion of very weak $D$ and $J_2$ terms can give rise to four different kinds of competing ground states [Fig.~\ref{fig1}(a)]. Therefore, at low temperatures the strong quantum fluctuations enhanced by $\mathcal{H}'$ (perturbations) of $\alpha$-CrOOH(D) may finally suppress the long-range 120$^o$ N\'{e}el order proposed in the ideal $S$ = 3/2 NN THAF model~\cite{Goetze2015Ground} (see below).

\begin{figure}[t]
\begin{center}
\includegraphics[width=8.6cm,angle=0]{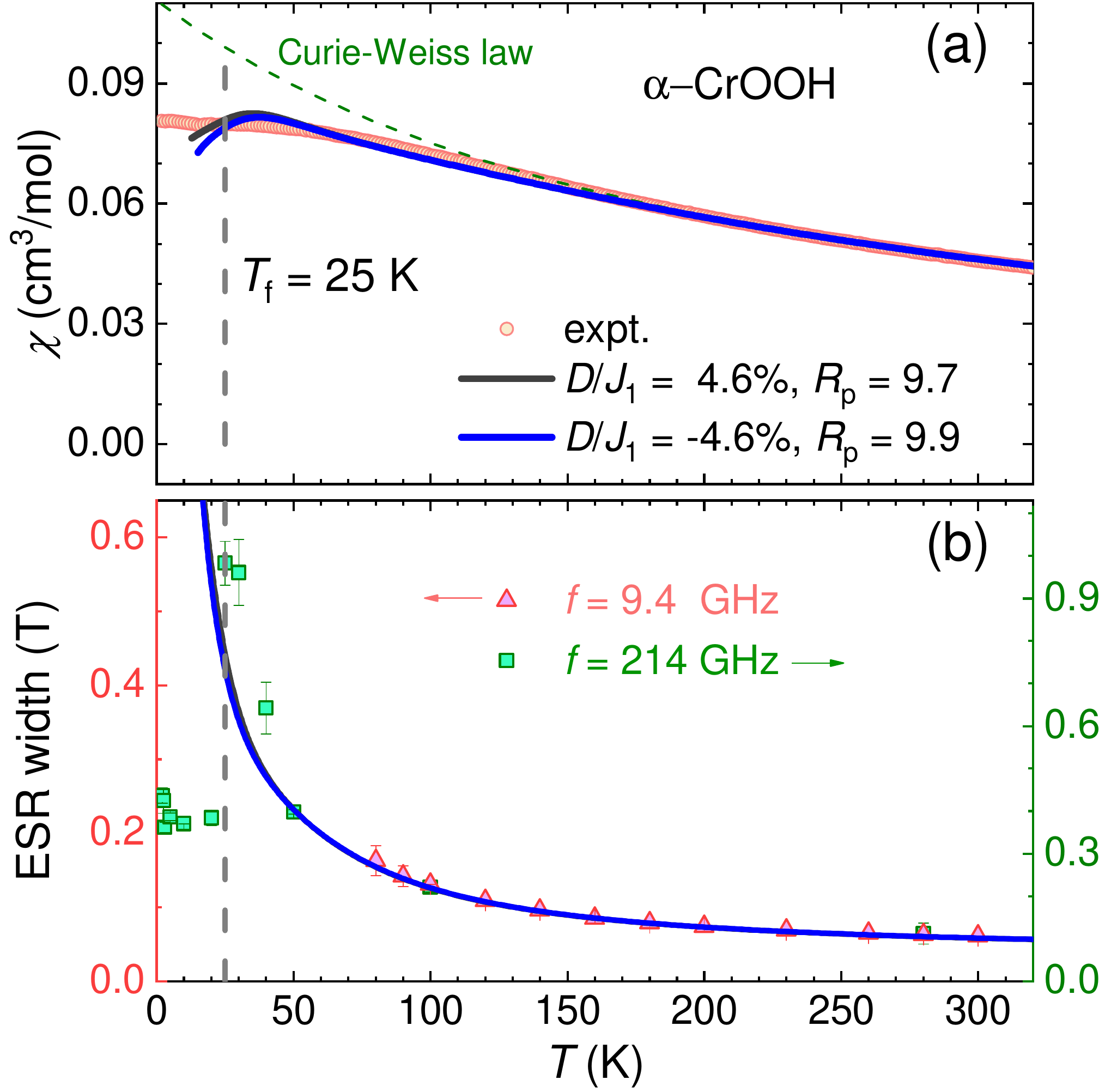}
\caption{(Color online)
Combined FLD fits to the temperature dependence of susceptibility (a) and X-band ($f$ = 9.4 GHz) ESR linewidth (b) measured on $\alpha$-CrOOH above 60 K, using the 12-site cluster with PBC. The dashed olive line in (a) represents the Curie-Weiss fit above 200 K. The high-field ESR linewidth measured at $f$ = 214 GHz is shown in (b) for comparison. The freezing temperature of $T_f$ = 25 K is marked by the dashed gray line in both (a) and (b).}
\label{fig2}
\end{center}
\end{figure}

Possibly owing to the absence of the DM interactions and strong coupling anisotropy, the paramagnetic ESR linewidth is relatively narrow at 300 K in $\alpha$-CrOOH, $\omega$ $\sim$ 0.0616(2) T at $f$ $\sim$ 9.4 GHz [Fig.~\ref{fig1}(e)], which is about an order of magnitude smaller than those reported in other QSL candidates, such as ZnCu$_3$(OH)$_6$Cl$_2$~\cite{PhysRevLett.101.026405} and YbMgGaO$_4$~\cite{li2015rare}. On the other hand, the small finite value should indicate there exists a weak single-ion anisotropy ($D$) in the spin Hamiltonian of $\alpha$-CrOOH(D) (see below for the quantitative analysis)~\cite{PhysRevB.4.38}. When the ESR frequency increases to $\sim$ 22.8 times, the linewidth is only slightly increased, $\omega$($f_2$=214 GHz) $\sim$ 1.7$\omega$($f_1$=9.4 GHz) [see Fig.~\ref{fig2}(b)], suggesting the $g$-factor randomness or anisotropy is tiny, $\Delta g$ $\sim$ $\mu_Bg^2$[$\omega$($f_2$)-$\omega$($f_1$)]/($hf_2$-$hf_1$) $\sim$ 0.006$g$. The contribution of $\Delta g$ to the ESR linewidth (Zeeman contribution) is minimized at $f$ $\sim$ 9.4 GHz, $\omega_\Delta$ = $\Delta ghf$/($\mu_Bg^2$) $\sim$ 2 mT $\ll$ $\omega$, and thus we can safely use the high-temperature X-band ESR linewidth to determine the single-ion anisotropy (see below).

The spin system of $\alpha$-CrOOH(D) is only weakly and almost linearly polarized in an applied magnetic field up to 14 T down to 2 K, $M$ $<$ 5\%$M_{sa}$, confirming the strong AF coupling [Fig.~\ref{fig1}(f)], where $M_{sa}$ = $gS$ $\mu_B$/Cr is the saturation magnetization. Because a high magnetic field of $\mu_0H_{1/3}$ $\sim$ 3$J_1S$/($g\mu_B$) $\sim$ 90 T is required to stabilize the 1/3 magnetization plateau, and the linear field dependence of magnetization is expected well below $\mu_0H_{1/3}$, according to the numerical result reported in the ideal Heisenberg case~\cite{Goetze2015Ground}. Through a linear fit to the magnetization above 3 T [Fig.~\ref{fig1}(f)], the concentration of magnetic moments of the impurities, $M_{im}$ $\leq$ 0.2\%$M_{sa}$, is obtained from the intercept, and is much smaller than that of many other frustrated magnets, e.g. 7.7\% in ZnCu$_3$(OH)$_6$Cl$_2$~\cite{PhysRevB.76.132411}. Furthermore, in $\alpha$-CrOOH(D) the magnetic moments of the impurities can be fully saturated at a low field of $\sim$ 1 T in the entire measured temperature range [see Fig.~\ref{figs2}(c)]~\footnote{Therefore, we measured the non-saturated intrinsic susceptibility of the spin system by $\chi$(3 T) = $\mu_0$[$M$(3.5 T)-$M$(2.5 T)] in this work.}.

\section{Fit to ESR linewidth and susceptibility}

High-temperature ESR linewidth is highly sensitive to the magnetic anisotropy of spin systems, and thus has been widely used in quantitative determination of the spin Hamiltonian parameters~\cite{PhysRevB.4.38,PhysRevLett.101.026405,li2015rare}. At $k_BT$ $\gg$ $g\mu_B\mu_0H$, the ESR linewidth is given by~\cite{PhysRevB.4.38},
\begin{equation}
\omega=\frac{\sqrt{2\pi}}{g\mu_B}\sqrt{\frac{M_2^3}{M_4}},
\label{Eq4}
\end{equation}
where $M_2$ = $\langle[\mathcal{H}',M^{+}][M^{-},\mathcal{H}']\rangle$/$\langle M^+M^-\rangle$ and $M_4$ = $\langle[\mathcal{H},[\mathcal{H}',M^{+}]][\mathcal{H},[\mathcal{H}',M^{-}]]\rangle$/$\langle M^+M^-\rangle$ are the second and fourth moments, respectively, $\langle\rangle$ represents a thermal average, and $M^{\pm}$ $\equiv$ $\Sigma_jS_j^{\pm}$ $\equiv$ $\Sigma_jS_j^{x}\pm i S_j^{y}$. Therefore, the perturbation terms in the Hamiltonian that don't commute with $M^{\pm}$ can contribute to the observed ESR linewidth, and we list below various origins commonly considered in the literature.

First, the hyperfine interactions broaden the ESR signal with $\omega_h$ $\sim$ $|A_h|^2$/($g\mu_BJ_1$) $<$ 0.3 $\mu$T, where $|A_h|$ $<$ 60 MHz is the hyperfine coupling of $^{53}$Cr$^{3+}$ (abundance of 9.55\%)~\cite{doi:10.1080/10420159508229826}. The dipole-dipole interactions also contribute to the linewidth with $\omega_d$ $\sim$ $|E_d|^2$/($g\mu_BJ_1$) $\sim$ 0.2 mT, where $|E_d|$ $\sim$ $\mu_0g^2\mu_B^2$/($4\pi a^3$) $\sim$ 1.9 GHz with the lattice parameter of $a$ = 2.98 {\AA}. There is only one Wyckoff position for Cr$^{3+}$ ions, and thus the almost uniform Zeeman interactions should not significantly contribute to the X-band ESR linewidth ($\omega_\Delta$ $\sim$ 2 mT, see above). All of the above contributions together account for a linewidth that is more than one order of magnitude smaller than the observed value ($\geq$ 0.06 T).

On the other hand, the single-ion anisotropy ($D$) terms in Eq.~(\ref{Eq3}) don't commute with $M^{\pm}$, and can give rise to the X-band ESR broadening. At $k_BT$ $\gg$ $J_1$$S^2$, $\langle\rangle$ $\rightarrow$ tr() is expected in Eq.~(\ref{Eq4}), and we obtain the ESR linewidth caused by $D$ ($J_2$ = 0 is fixed) as,
\begin{equation}
\omega(T\rightarrow\infty)=\frac{\sqrt{2\pi}}{g\mu_B}\sqrt{\frac{2.4^3D^4}{108J_1^2+9.6D^2}}.
\label{Eq5}
\end{equation}
Taking $J_1$ $\sim$ 28 K and $\omega$(300 K) $\sim$ 0.06 T (see above), we reach a rough estimation of $|D|$ $\sim$ 1.6 K.

Below $\sim$ 150 K, the measured susceptibility obviously deviates from the Curie-Weiss behavior [see Fig.~\ref{fig2}(a)] and the X-band ESR linewidth clearly depends on temperature [see Fig.~\ref{fig2}(b)], indicating the formation of low-$T$ AF correlations. Therefore, we perform many-body calculations to better simulate these finite-temperature observables, and to determine the Hamiltonian parameters more precisely. The sample difference of thermodynamic property between $\alpha$-CrOOH and $\alpha$-CrOOD gets significant below $\sim$ 60 K [see Figs.~\ref{figs2}(a) and \ref{figs2}(b)], and thus we apply the FLD calculation~\cite{Lanczos1950An,PhysRevB.49.5065} to fit the observables only above 60 K, where the size (12 sites) effect of FLD is also negligible (Appendix~\ref{a2}). First of all, we perform a combined fit to both susceptibility and X-band ESR linewidth above 60 K by fixing $J_2$ = 0 [Fig.~\ref{fig2}], and obtain $J_1$ = 23.5 K and $D$ = 1.08 K with the least-$R_p$ = 9.7. In the easy-axis region ($D$ $<$ 0), we find another minimum-$R_p$ point in the parameter space, $J_1$ = 23.6 K and $D$ = -1.08 K with $R_p$ = 9.9. Furthermore, we also try to determine $J_2$. Unfortunately, the quality of the fit is not significantly improved with a non-zero $J_2$, as both susceptibility and ESR linewidth~\footnote{The $J_2$ terms commute with $M^{\pm}$.} are insensitive to $J_2$ above 60 K.

\begin{figure}[t]
\begin{center}
\includegraphics[width=8.2cm,angle=0]{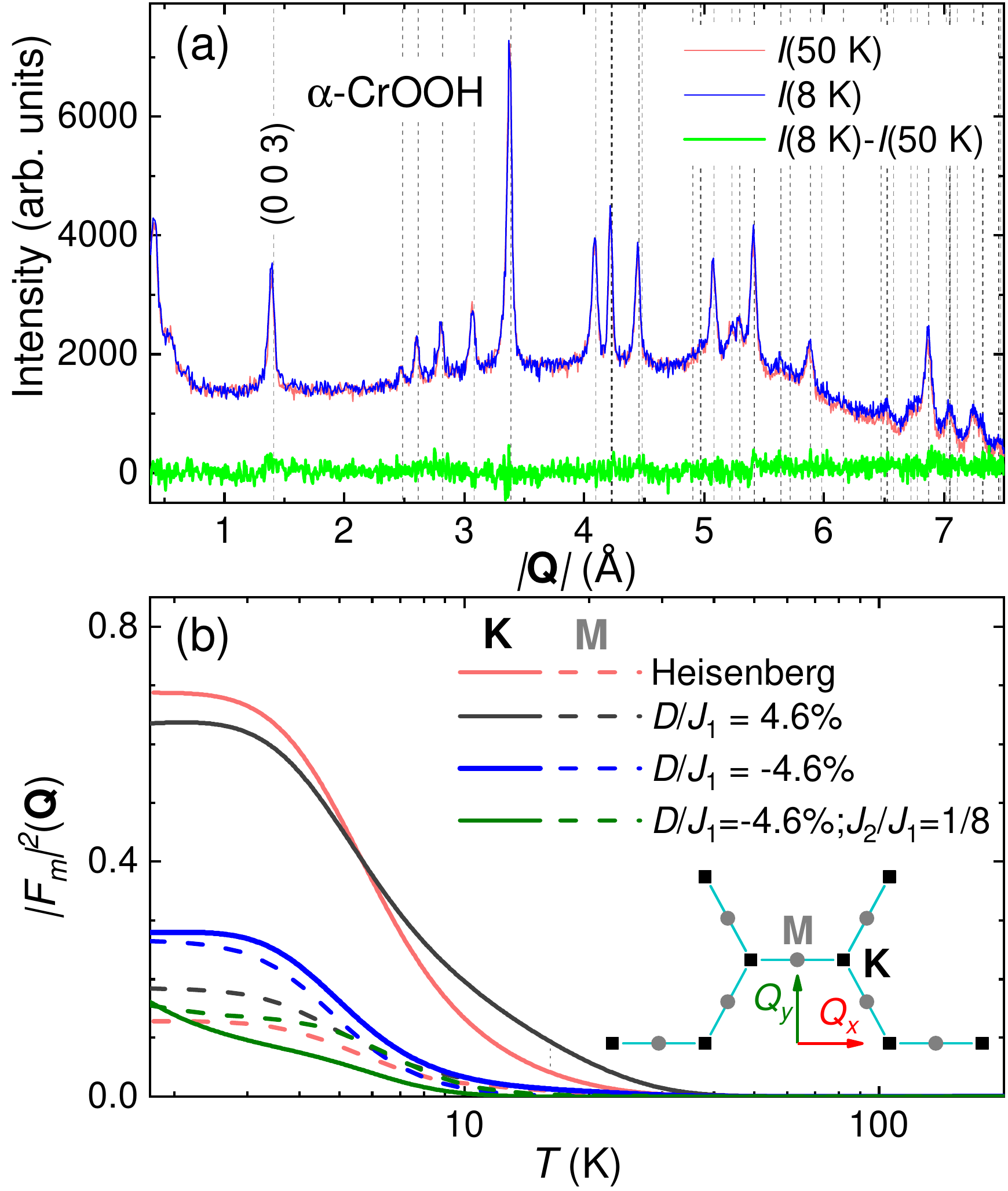}
\caption{(Color online)
(a) Neutron powder diffraction patterns of $\alpha$-CrOOH measured at 8 K [$I$(8 K)] and 50 K [$I$(50 K)] with the green line showing the difference, $I$(8 K)-$I$(50 K). The nuclear reflections are marked by the dashed lines. (b) Temperature dependence of static structure factor at the K (solid) and M (dashed) points calculated by the 12-site FLD using different sets of Hamiltonian parameters ($J_1$ = 23.5 K). The inset shows Brillouin zone boundaries, with the wave vectors of the three-sublattice (K) and stripe (M) orders marked by black squares and gray dots, respectively.}
\label{fig3}
\end{center}
\end{figure}

\section{Absence of long-range spin order}

\begin{figure}[t]
\begin{center}
\includegraphics[width=8.2cm,angle=0]{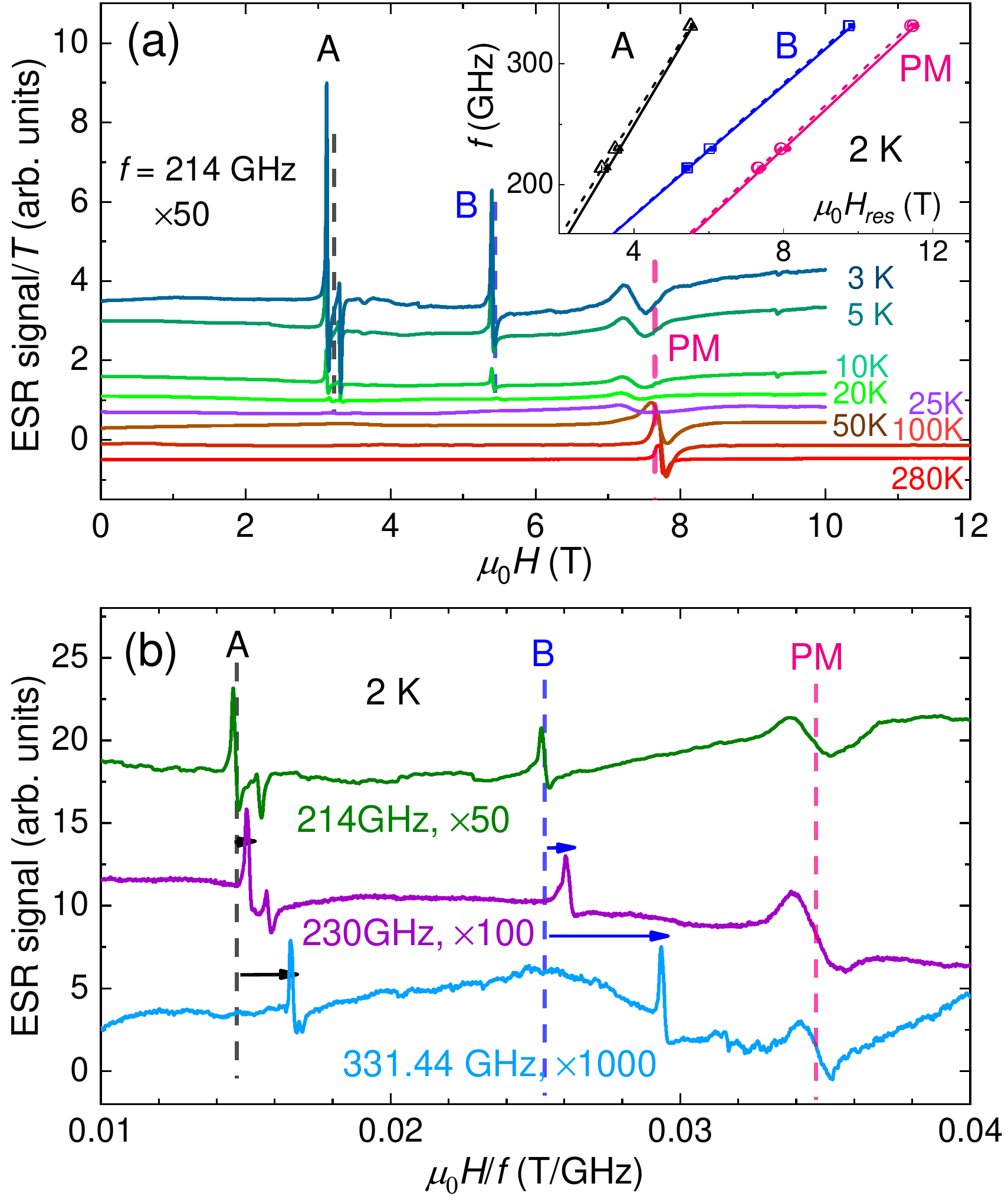}
\caption{(Color online)
(a) High-field ESR spectra (divided by temperature) of $\alpha$-CrOOH measured with $f$ = 214 GHz at various temperatures. Three different resonance modes observed below $T_f$ = 25 K are marked by A, B, and PM, respectively. The inset shows the frequency dependence of resonance fields of three modes measured at 2 K. The open and solid scatters represent the resonance fields along the $ab$ plane and $c$ axis, respectively, with the dashed and solid lines showing the corresponding linear fits. (b) ESR signals measured with selected frequencies at 2 K. The field shifts of the additional modes (A and B) are marked by arrows. The first derivative absorption spectra are vertically shifted for a better view in both (a) and (b).}
\label{fig4}
\end{center}
\end{figure}

We do observe a weak anomaly (kink) in heat capacity at $T_f$ = 25 K for $\alpha$-CrOOH or at $T_f$ = 22 K for $\alpha$-CrOOD [Fig.~\ref{figs2}(b)], as reported previously~\cite{Matsuo2006Isotope,maekawa2007deuteration-induced}. While, the absence of long-range magnetic order in $\alpha$-CrOOH(D) is evidenced by the following experimental observations. (\romannumeral1) No convincing magnetic reflection in neutron diffraction is observed at 8 K, as there is no obvious difference between the neutron diffraction patterns measured at 8 and 50 K [see Figs.~\ref{fig3}(a) and \ref{figs1}(b)]. For $\alpha$-CrOOH, the upper bound of the static structure factor at $|\mathbf{Q}_K|$ = 4$\pi$/(3$a$) is roughly estimated as, $\langle|$$I$(8 K)-$I$(50 K)$|\rangle$/[6$\times$(5.4 fm)$^2S_{ph}|f(|\mathbf{Q}_K|)|^2$] $<$ 0.2 per Cr~\cite{PhysRevX.10.011007}, that is much smaller than the numerical result reported for the 120$^o$ N\'{e}el order, $M_{sub}^2$/2 $\sim$ 0.75 per site, on the ideal $S$ = 3/2 NN THAF model~\cite{Goetze2015Ground}. Here, $S_{ph}$ = $I_n$/($2|F_n|^2$) $\sim$ 11.9 fm$^{-2}$Cr is the scale factor with $I_n$ = 2160(140) and $|F_n|^2$ = 91 fm$^2$/Cr presenting the observed intensity and calculated structure factor of the (0 0 3) nuclear reflection [see Fig.~\ref{fig3}(a)], respectively, and $\langle|$$I$(8 K)-$I$(50 K)$|\rangle$ $<$ 300. In contrast, clear magnetic reflections or humps in the neutron powder diffraction patterns were observed in CuCrO$_2$, AgCrO$_2$, Cu$_{0.85}$Ag$_{0.15}$CrO$_2$, and CuCr$_{1-y}$Al$_{y}$O$_2$ (0 $<$ $y$ $\leq$ 0.2) around their $n|\mathbf{Q}_K|$ ($n$ = 1, 2, ...) at low temperatures~\cite{Kadowaki1990Neutron,Okuda2009Dimensional,doi:10.1143/JPSJ.80.014711}. Our observation is also obviously different from that reported in NiGa$_2$S$_4$, where the strongest magnetic peak appears at $|\mathbf{Q}|$ of ($\frac{1}{6}$ $\frac{1}{6}$ 0)~\cite{Nakatsuji2005Spin} and the antiferro nematic order may form at low temperatures~\cite{doi:10.1143/JPSJ.75.083701}. (\romannumeral2) The broad paramagnetic ESR signal with dominant integral intensity maintains clearly visible down to 2 K [see Fig.~\ref{fig4}]. (\romannumeral3) The spin excitations of $\alpha$-CrOOD measured by INS keep continuous down to 3.5 K, whereas the dispersion of the 120$^o$ N\'{e}el order gets very sharp in the powder spectra calculated using the ideal THAF model at low temperatures (see Fig.~\ref{fig7}). (\romannumeral4) No sharp $\lambda$ peak is observed in the specific heat of $\alpha$-CrOOH or $\alpha$-CrOOD at $T_f$ [see Fig.~\ref{figs2} (b)]. In sharp contrast, the temperature dependence of specific heat of CuCrO$_2$ shows a sharp $\lambda$ peak at the N\'{e}el temperature, $T_N$ $\sim$ 25.8 K~\cite{PhysRevB.77.134423}.

The single-ion easy-axis anisotropy, $D$ $<$ 0, in Eq.~(\ref{Eq3}), makes it energetically favourable for the spins to align along the $c$ axis, and thus can induce frustration/competition to the 120$^o$ N\'{e}el order. Compared to the ideal $S$ = 3/2 NN THAF model, the inclusion of the easy-axis anisotropy of $D$/$J_1$ = -4.6\% can significantly suppress the 120$^o$ N\'{e}el order, as evidenced by the decrease of the calculated static structure factor at the K points [see Fig.~\ref{fig3}(b)]. Similarly, a small NNN Heisenberg AF coupling, $J_2$/$J_1$ $\sim$ 1/8, can further frustrate the three-sublattice magnetic structure at low temperatures [see Fig.~\ref{fig3}(b)]~\cite{doi:10.7566/JPSJ.83.093707,PhysRevB.92.041105,PhysRevB.91.014426,PhysRevLett.123.207203}. Therefore, we argue that the single-ion anisotropy and (or) NNN interactions play an important role to stabilize the possible QSL ground state in $\alpha$-CrOOH(D) [see Fig.~\ref{fig1}(a)]. Moreover, we also notice that the measured thermodynamic data of $\alpha$-CrOOD slightly deviates from that of $\alpha$-CrOOH below $\sim$ 60 K [see Figs.~\ref{figs2}(a) and \ref{figs2}(b)]. The replacement of H by heavier D should mainly adjust the lattice vibrations, and thus may also slightly modify the low-temperature magnetism of the system through spin-lattice couplings.

\section{High-field magnetic resonance}

\begin{figure}[t]
\begin{center}
\includegraphics[width=8.7cm,angle=0]{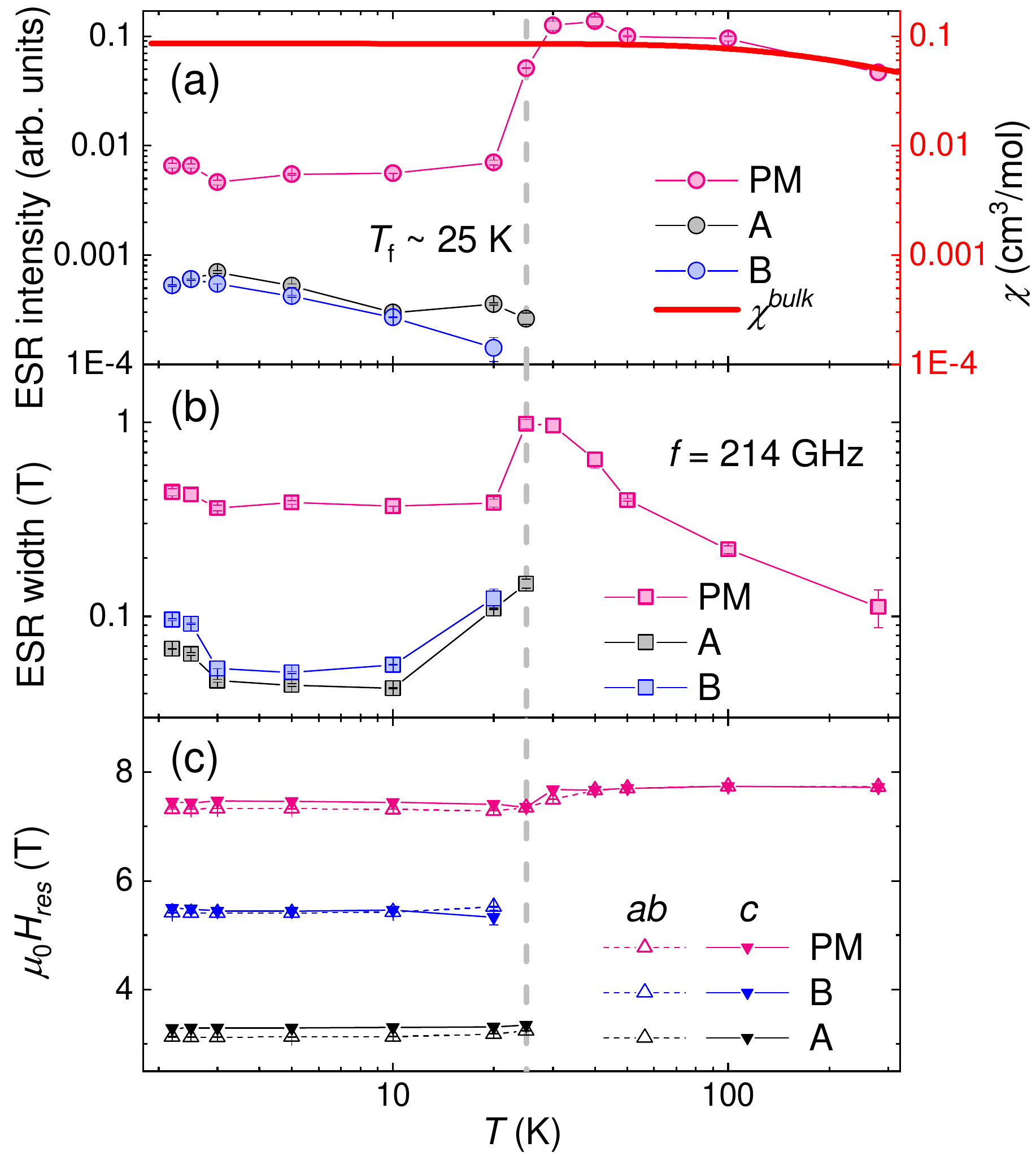}
\caption{(Color online)
Temperature dependence of ESR (a) integral intensity, (b) linewidth, and (c) resonance field of three modes measured at $f$ = 214 GHz on $\alpha$-CrOOH. The bulk susceptibility is shown in (a) for comparison. The freezing temperature of $T_f$ = 25 K is marked by the dashed gray line. In (c), the open and solid scatters show the resonance fields along the $ab$ plane and $c$ axis, respectively.}
\label{fig5}
\end{center}
\end{figure}

High-field and high-frequency magnetic resonance technique is very powerful in separating magnetic resonating signals with different origins. Beside the broad paramagnetic signal (PM), two narrow additional modes (A and B) are observed below $T_f$ = 25 K in $\alpha$-CrOOH at $f$ = 214 GHz [see Fig.~\ref{fig4}(a)]. To investigate the origin of these additional ESR signals, we performed more ESR measurements at higher frequencies and 2 K, and presented them together after dividing the sweeping field by the frequency [see Fig.~\ref{fig4}(b)]. In contrast to the broad paramagnetic mode, both additional ESR signals of $\mu_0H_{res}$/$f$ obviously shift to higher fields with increasing the frequency, thus precluding the paramagnetic origin. The frequency dependence of the resonance fields of both modes is almost linear [see inset of Fig.~\ref{fig4}(a)], and can be described empirically using the linear (Kittel) equation~\cite{PhysRevLett.112.077206},
\begin{equation}
hf=g_{m,\eta}\mu_B\mu_0(H_{res}^{\eta}+H_{m,\eta}^a),
\label{Eq6}
\end{equation}
where $g_{m,\eta}$ is the effective $g$ factor, $\mu_0H_{m,\eta}^a$ is the anisotropy field, $m$ = A, B, PM and $\eta$ = $ab$, $c$ are the mode and orientation indexes, respectively. The fitted $g_{m,\eta}$ and $\mu_0H_{m,\eta}^a$ are summarized in Table~\ref{table1}. The fitted non-zero anisotropy fields of both A and B modes correspond to very weak correlation energies of $\sim$ $g_{m,\eta}\mu_B\mu_0H_{m,\eta}^a$ $\sim$ 0.1$J_1$, in the framework of spin-wave theory~\cite{PhysRevLett.112.077206}. Moreover, the integral intensities of both A and B modes are only about an order of magnitude smaller than that of the main PM mode [see Fig.~\ref{fig5}(a)]. The above observations suggest that a small amount ($\sim$ 10\%) of electronic spins freezes into some kind of very weak (ferromagnetic or antiferromagnetic) order, while the majority remains dynamic below $T_f$ in $\alpha$-CrOOH.

\begin{table}[t]
\caption{Anisotropy fields (unit: T) and effective $g$ factors obtained by linear fits to the magnetic resonance fields measured at 2 K [inset of Fig.~\ref{fig4}(a)].}\label{table1}
\begin{center}
\begin{tabular}{ l || l | l || l | l || l | l }
    \hline
    \hline
 $m$,$\eta$ & A,$ab$ & A,$c$ & B,$ab$ & B,$c$ & PM,$ab$ & PM,$c$ \\ \hline
 $\mu_0H_{m,\eta}^a$ & 1.17(5) & 0.86(3) & 2.46(3) & 2.5(1) & 0 & 0 \\ \hline
 $g_{m,\eta}$ & 3.55(3) & 3.68(2) & 1.938(6) & 1.92(2) & 2.082(4) & 2.054(4) \\
    \hline
    \hline
\end{tabular}
\end{center}
\end{table}

The ESR intensity of the PM mode is proportional to the imaginary part of the dynamic susceptibility, $\chi"$(\textbf{Q}$\rightarrow$0, $E$), and roughly follows the bulk susceptibility above $T_f$ $\sim$ 25 K [see Fig.~\ref{fig5}(a)]. While, below $\sim$ $T_f$ the paramagnetic intensity obviously deviates from the bulk susceptibility, and is quickly suppressed to a very small constant, $\chi_0^{PM}$ $\sim$ 0.006(1) cm$^3$/mol, possibly suggesting the formation of a \emph{spinon} Fermi pocket in the resonance field of $\sim$ 7.5 T $>$ $k_BT$/$\mu_B$~\cite{PhysRevLett.98.117205}, where $T$ $\sim$ 2 K (see below). In contrast, the dc bulk susceptibility is static including the contributions from all three modes, and reaches to a much larger constant in the zero-temperature limit, $\chi_0^{bulk}$ $\sim$ 0.08 cm$^3$/mol [see Fig.~\ref{figs2}(a)]. Around $T_f$, the ESR linewidths of all three modes take maximum values [Fig.~\ref{fig5}(b)], possibly suggesting the starting of long-range spin-spin entanglements or intrinsic topological orders in the spin system of $\alpha$-CrOOH. Moreover, below $\sim$ $T_f$ a tiny anisotropy of the resonance fields develops [Fig.~\ref{fig5}(c)], and the paramagnetic resonance fields weakly decrease, corresponding to an increase of the $g$ factors from $g_{ab}$ $\sim$ $g_{c}$ $\sim$ 1.97 (above $T_f$) to $g_{ab}$ $\sim$ 2.08 and $g_{c}$ $\sim$ 2.05 (below $T_f$).

\begin{figure}[t]
\begin{center}
\includegraphics[width=8.0cm,angle=0]{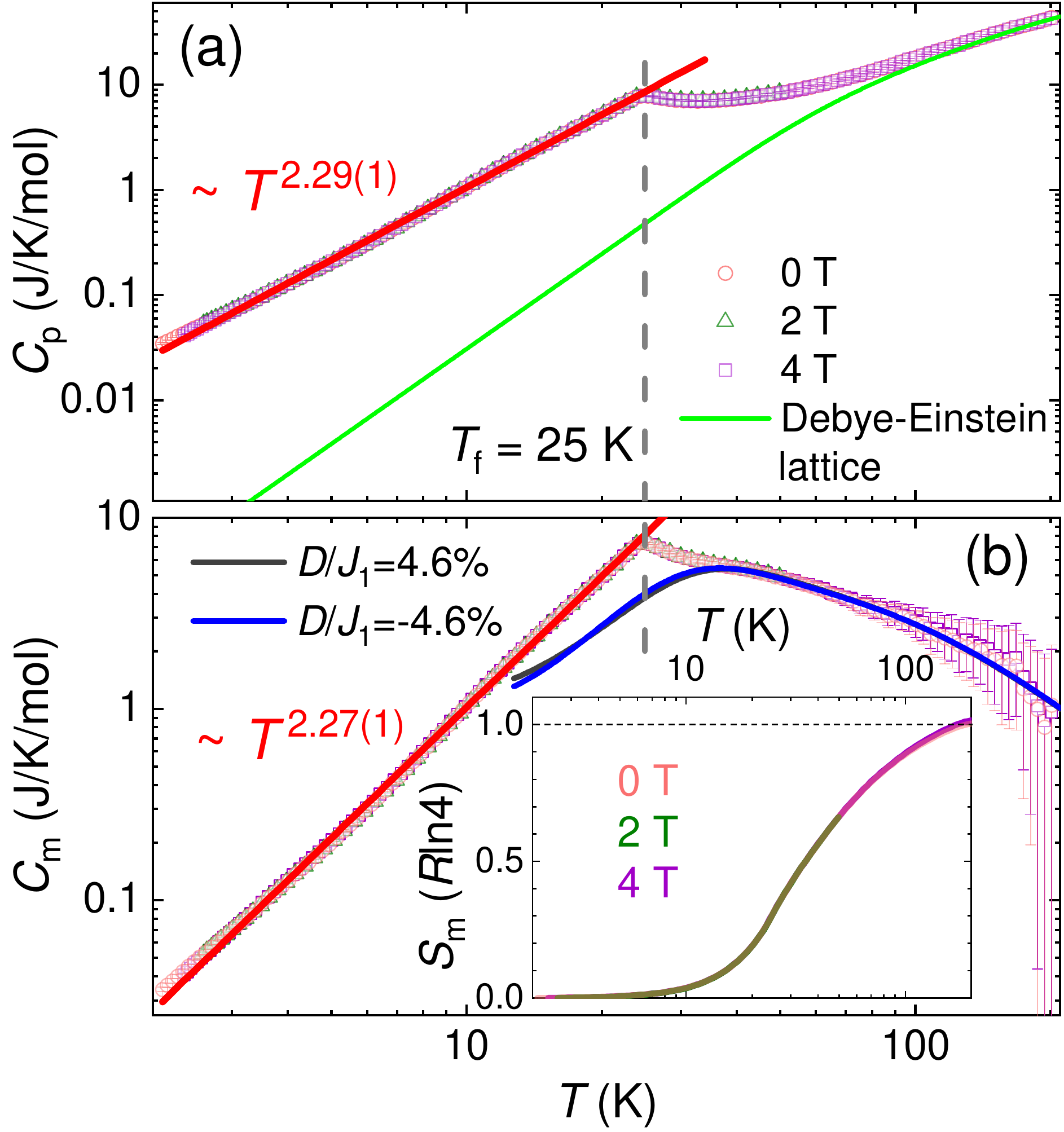}
\caption{(Color online)
(a) Temperature dependence of specific heat measured on $\alpha$-CrOOH at 0, 2, and 4 T, respectively. The green line presents the lattice contribution. (b) Temperature dependence of magnetic heat capacity. The  black and blue lines show the magnetic contributions calculated using two different sets of parameters (see main text). The inset presents the magnetic entropy obtained by integrating $C_m$/$T$ from $T$ = 2 K. The power-law behaviors of specific heat are shown by red lines below $T_f$ = 25 K in both (a) and (b).}
\label{fig6}
\end{center}
\end{figure}

\section{Power-law behavior of specific heat}

\begin{figure*}[t]
\begin{center}
\includegraphics[width=18cm,angle=0]{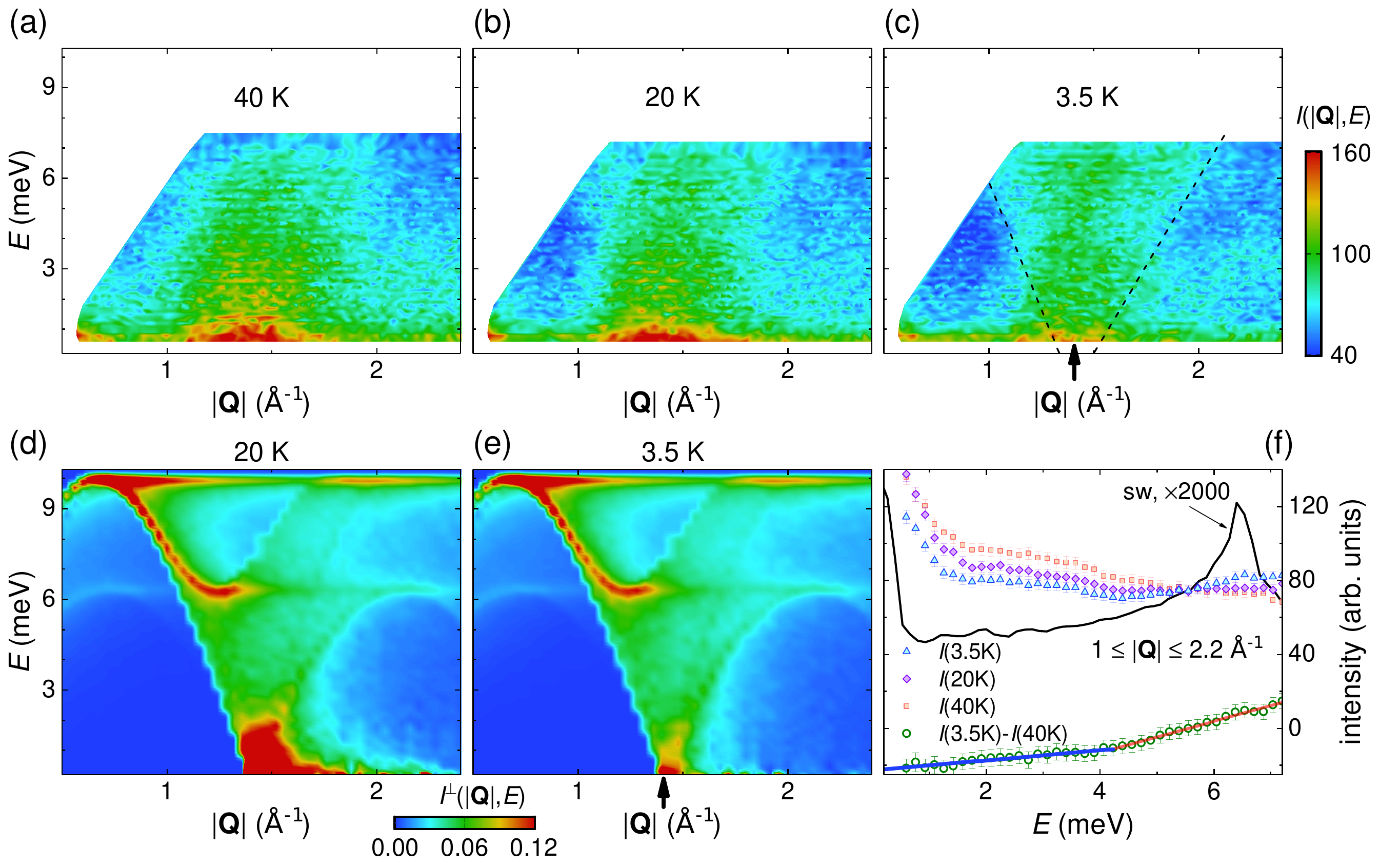}
\caption{(Color online)
INS spectra measured on the powder of $\alpha$-CrOOD at (a) 40, (b) 20, and (c) 3.5 K, respectively. Powder spectra of the spin-wave excitations calculated for the ideal $S$ = 3/2 THAF model with $J_1$ = 2.1 meV at (d) 20 and (e) 3.5 K, respectively, based on the linear spin-wave theory. The arrows present the momentums at the K points, $|\mathbf{Q}|$ = 4$\pi$/(3$a$), where the Dirac cones [roughly marked by the dashed lines in (c)] emerge. (f) Energy dependence of the INS intensities integrated over the momentum space, 1 $\leq$ $|\mathbf{Q}|$ $\leq$ 2.2 {\AA}$^{-1}$. The blue and red lines present the linear fits to the data [$I$(3.5K)-$I$(40K)] at 0.6 $\leq$ $E$ $\leq$ 4.2 meV and 4.2 $\leq$ $E$ $\leq$ 7.2 meV, respectively, and the black line shows the calculated spin-wave dependence at 3.5 K for comparison.}
\label{fig7}
\end{center}
\end{figure*}

Above $\sim$ 60 K, the magnetic specific heat of $\alpha$-CrOOH can be precisely calculated using the TAF model with the fitted parameters of $J_1$ $\sim$ 23.5 K and $|D|$ $\sim$ 1.08 K ($J_2$ = 0), which has successfully simulated the measured bulk susceptibility and ESR linewidth (see Fig.~\ref{fig2}). Therefore, at first we obtain the precise lattice specific heat of $\alpha$-CrOOH by subtracting the calculated magnetic contribution [the black and blue lines in Fig.~\ref{fig6}(b)] from the total measured specific heat, above 60 K [see Fig.~\ref{figs2}(d)]. Then, we get the precise Debye-Einstein specific heat in the entire temperature range by fitting to the above lattice specific heat between 60 and 200 K [see Appendix~\ref{a1}]. After subtracting the Debye-Einstein part [the green line in Fig.~\ref{fig6}(a)], we obtain the magnetic specific heat, $C_m$, and then the magnetic entropy ($S_m$) by a integral [see Fig.~\ref{fig6}(b)]. From 2 to 200 K, the increase of the magnetic entropy, $\Delta S_m$ = 1.004(12)$R$ln4 $\sim$ $R$ln(2$S$+1), precisely obeys the third law of thermodynamics, and justifies the above data processing. Below 10 K, the temperature dependence of magnetic entropy flattens out with $S_m$ $\leq$ 3.5\%$R$ln4, suggesting the ground-state property of the spin system is nearly approached.

Below $T_f$, the magnetic specific heat of $\alpha$-CrOOH shows a power-law temperature dependence, $C_m$ $\sim$ $\beta T^{\gamma}$, with $\gamma$ = 2.27(1) and $\beta$ = 0.0055(1) JK$^{-1-\gamma}$mol$^{-1}$ [see Fig.~\ref{fig6}(b)]. This power-law behavior is further confirmed by the direct analysis on the total specific heat ($C_p$) measured on both $\alpha$-CrOOH and $\alpha$-CrOOD under various applied magnetic fields, as the lattice contribution is more than one order of magnitude smaller than $C_p$ below $T_f$ [see Fig.~\ref{fig6}(a)]. In $\alpha$-CrOOD, the similar power-law temperature dependence of $C_p$ is observed as well, with $\gamma$ = 2.166(2) and $\beta$ = 0.00845(4) JK$^{-1-\gamma}$mol$^{-1}$ [see Fig.~\ref{figs2}(b)]. Therefore, we perform an average, and get $C_m$ $\sim$ 0.007$T^{2.2}$ JK$^{-1}$mol$^{-1}$ for $\alpha$-CrOOH(D). Similar to NiGa$_2$S$_4$~\cite{Nakatsuji2005Spin}, a very low sensitivity of specific heat to the external magnetic field is observed in both $\alpha$-CrOOD (Fig.~\ref{fig6}) and $\alpha$-CrOOH [Fig.~\ref{figs2}(b)], which is well consistent with the nearly linear field dependence of magnetization ($M$) and thus the nearly field-independent susceptibility ($dM$/$dH$) measured up to 14 T [see Fig.~\ref{fig1}(f)]. This effect is a very strong signature of the presence of nonmagnetic singlet-like~\footnote{The ``valence bonds'' in spin-3/2 TAF systems are more complicated than the singlets in the spin-1/2 cases.} excitations and should not be expected in a conventional N\'{e}el-ordered or spin-glass system~\cite{PhysRevLett.84.2953}. The above observations consistently verify the intrinsic power-law behavior of specific heat in the spin-3/2 system of $\alpha$-CrOOH(D).

\section{Gapless Dirac cones}

The low-energy-transfer INS measurements were performed both above and below the freezing temperature of $T_f$ = 22 K for $\alpha$-CrOOD, and a continuum of the excitations broadly distributed in both momentum ($|\mathbf{Q}|$) and energy ($E$) space is clearly visible [see Figs.~\ref{fig7}(a)-\ref{fig7}(c)]. The spectral weight of the continuum is mainly distributed at low momenta, thus largely excluding the phonon origin of the excitations~\cite{Kajimoto2018Elastic}. The continuum centered around $|\mathbf{Q}|$ = 4$\pi$/(3$a$) becomes well cone-shaped below $T_f$ = 22 K [see Fig.~\ref{fig7}(c)]~\footnote{The asymmetry of the cone-shape signal should be caused by the powder average of spectra in the wave-vector space.}, and should be related to the putative Dirac nodes of the gapless \emph{U}(1) QSL at the K points of the Brillouin zone [see inset of Fig.~\ref{fig3}(b)]~\cite{PhysRevLett.123.207203}. Recently, the very similar spectral weight accumulating of the low-energy continuum at the K points was also reported on the $S$ = 1/2 TAF powder of NaYbO$_2$ with strong spin-orbit couplings~\cite{PhysRevB.100.144432}. At low temperatures, the calculated spin-wave excitations of the 120$^o$ N\'{e}el order on the ideal $S$ = 3/2 THAF model show very sharp dispersion, as well as strong elastic Bragg peaks with a direct band gap~\cite{li2017nearest} of $\sim$ 3$J_1$ at $|\mathbf{Q}|$ = 4$\pi$/(3$a$), 8$\pi$/(3$a$), etc. [see Figs.~\ref{fig7}(e) and \ref{fig7}(f)], and thus obviously can't account for the excitation continuum observed in $\alpha$-CrOOD. The difference of the INS intensities [$I$(3.5K)-$I$(40K)] roughly measuring the density of states shows nearly linear energy dependence at 3.5 K [see Fig.~\ref{fig7}(f)], possibly further confirming the formation of the well-defined Dirac cones at the K points in $\alpha$-CrOOD~\cite{PhysRevLett.102.047205,Biswas2011explo}.

\section{Ground state and discussion}

Although the reports of spin liquids in a $S$ = 3/2 TAF system are rare, the possibility of the formation of a gapless \emph{U}(1) Dirac spin-liquid ground state in the $S$ = 1/2 or 1 system on triangular or kagome lattice has been extensively studied theoretically~\cite{PhysRevLett.123.207203,PhysRevB.81.224417,PhysRevB.93.144411,PhysRevLett.84.2953,PhysRevLett.98.117205,PhysRevLett.102.047205,PhysRevX.7.031020,CHEN20181545}. Owing to the ideal Dirac nodes, the power-law behavior of specific heat, $C_m$ $\propto$ $T^2$, is expected, and our observation of $C_m$ $\propto$ $T^{2.2}$ in $\alpha$-CrOOH(D) (Fig.~\ref{fig6}) is well consistent with this prediction. On the other hand, the ideal Dirac QSL is also predicted to give rise to a linear temperature dependence of susceptibility at 0 T~\cite{PhysRevB.81.224417}, which seems in contrast to that of the paramagnetic ESR intensity observed in $\alpha$-CrOOH at low temperatures [Fig.~\ref{fig5}(a)]. In the presence of various perturbations in the real material of $\alpha$-CrOOH, the ESR intensity indeed decreases by about an order of magnitude from $T_f$ = 25 K down to 2 K, which may be not far away from the predicted linear behavior in the ideal case at 0 T. On the other hand, the PM mode is measured in a finite (resonance) magnetic field of $\sim$ 7.5 T, the putative \emph{spinon}s will form a Fermi pocket with a radius proportional to the field strength~\cite{PhysRevLett.98.117205}, and thus the small finite constant of the ESR intensity observed at low temperatures [Fig.~\ref{fig5}(a)] is well understandable in this framework. Moreover, the recent numerical result of $C_m$ $\propto$ $T^2$ and $\chi$ $\sim$ \emph{constant} reported in the gapless algebraic paramagnetic liquid~\cite{CHEN20181545} may provide another possible resolution to this contradiction. The signature of the Dirac cones observed at the K points in the low-energy continuum of $\alpha$-CrOOD [see Fig.~\ref{fig7}(c)] further verifies the above explanation.

The finite spin susceptibility and power-law behavior of specific heat observed at low temperatures in $\alpha$-CrOOH(D) may be explained by other gapless spin-liquid ground states, e.g. the spin nematic phase, previously proposed to understand the frustrated magnetism of the $S$ = 1 triangular magnet, NiGa$_2$S$_4$~\cite{Nakatsuji2005Spin,doi:10.1143/JPSJ.75.083701,PhysRevB.74.092406}. While, the biqudratic interactions that are generally present in spin-1 magnets and favor nematic ordering may be absent or tiny in the $S$ = 3/2 magnet of $\alpha$-CrOOH(D). Furthermore, the absence of any convincing magnetic reflections in neutron diffraction of $\alpha$-CrOOH(D) seems to speak against this antiferro nematic order with a three-sublattice structure~\cite{doi:10.1143/JPSJ.75.083701}.

\section{Conclusions}

We study the frustrated magnetism of the $S$ = 3/2 TAF $\alpha$-CrOOH(D) (delafossites green-grey powder), by neutron scattering, ESR, and thermodynamic measurements. The nearly Heisenberg Hamiltonian with a weak single-ion anisotropy ($J_1$ $\sim$ 23.5 K and $|D|$/$J_1$ $\sim$ 4.6\%) is quantitatively determined by fitting to the temperature dependence of ESR linewidth and bulk susceptibility measured above 60 K. Owing to the single-ion anisotropy and (or) other perturbations, the spin system of $\alpha$-CrOOH(D) keeps disordered with clear paramagnetic resonance signal down to 2 K ($\sim$ 0.04$J_1$$S^2$), where the magnetic entropy is almost zero. The power-law temperature dependence of specific heat observed below $T_f$ is insensitive to the external magnetic field, consistent with the theoretical prediction of a gapless QSL ground state with Dirac nodes. This is further confirmed by the significantly suppressing of the paramagnetic resonance signal and by the direct observation of the signature of the Dirac nodes at the K points of the Brillouin zone, below $\sim$ $T_f$. Our work may shed new light on the search for QSLs in frustrated systems with $S$ = 3/2.

\acknowledgements

We thank Zhongwen Ouyang and Lei Yin for ESR measurements, as well as Qingming Zhang, Shun Wang, Jianshu Li, Feng Jin, and Quanwen Zhao for specific heat measurements. Y.L. thanks Bin Xi for helpful discussion on the many-body simulation. W.T. was supported by the National Key Research and Development Program of China (Grant No. 2017YFA0403502) and the National Natural Science Foundation of China (No. U1732275). This work was supported by the National Natural Science Foundation of China (No. 11875238) and the Fundamental Research Funds for the Central Universities, HUST: 2020kfyXJJS054.

\begin{figure}[t]
\begin{center}
\includegraphics[width=8.0cm,angle=0]{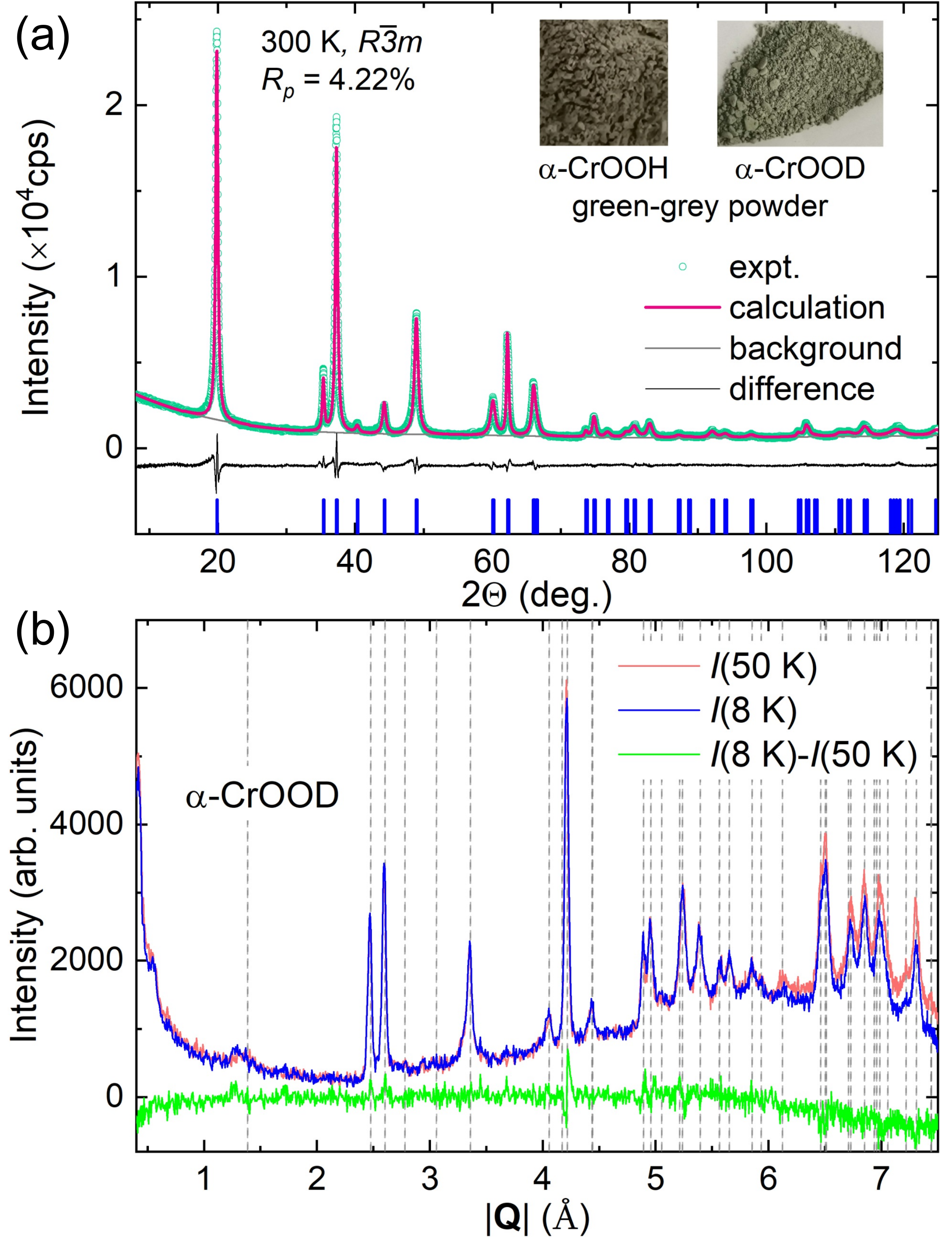}
\caption{(Color online)
(a) X-ray powder diffraction measured at 300 K and Rietveld refinement for $\alpha$-CrOOH. The insets show the green-grey powders of $\alpha$-CrOOH and $\alpha$-CrOOD. (b) Neutron diffraction spectra of $\alpha$-CrOOD measured at 8 K [$I$(8 K)] and 50 K [$I$(50 K)], respectively. The green line shows the difference [$I$(8 K)-$I$(50 K)], and the dashed grey lines present the nuclear reflections.}
\label{figs1}
\end{center}
\end{figure}

\section*{Appendix}

\appendix

\section{Sample characterization.}
\label{a1}

\begin{figure}[t]
\begin{center}
\includegraphics[width=8.7cm,angle=0]{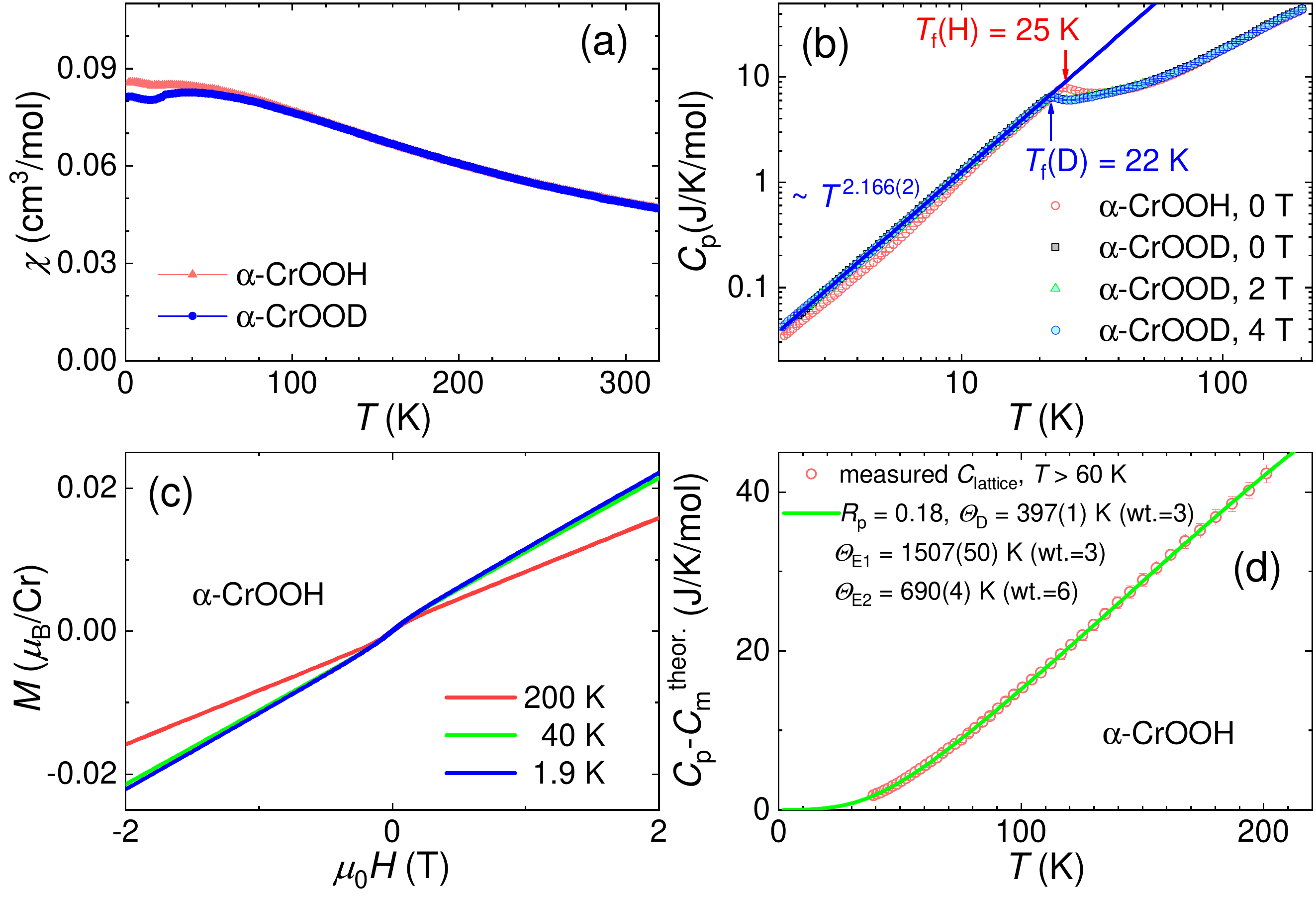}
\caption{(Color online)
Temperature dependence of (a) bulk susceptibility measured at 3 T and (b) specific heat measured at selected fields on $\alpha$-CrOOD together with the corresponding data measured on $\alpha$-CrOOH. The freezing temperatures of $\alpha$-CrOOD (22 K) and $\alpha$-CrOOH (25 K) are marked. (c) Complete magnetization loops measured at selected temperatures for $\alpha$-CrOOH. (d) Lattice specific heat of $\alpha$-CrOOH obtained by subtracting the calculated magnetic contribution ($C_m^{theor.}$, see main text) from the measured (total) specific heat [see (b)]. The green line shows the least-$R_p$ fit above 60 K using the combination of the Debye and Einstein functions [Eq.~(\ref{eqs1})], and three fitted parameters are given.}
\label{figs2}
\end{center}
\end{figure}

The powders of $\alpha$-CrOOH and $\alpha$-CrOOD are shown in Fig.~\ref{figs1} (a). Both x-ray and neutron powder diffraction measurements preclude there exist any significant impurity phases ($<$ 5\%) or structure transitions, as no additional peak is observed at least down to 8 K in $\alpha$-CrOOH(D) (see Fig.~\ref{figs1}). The results of the structure refinements are summarized in Table~\ref{table2}. The sample difference of thermodynamic property between $\alpha$-CrOOH and $\alpha$-CrOOD is neglectable above $\sim$ 60 K, but becomes significant below $\sim$ 60 K. While, the overall temperature dependence behaviors are still largely identical [see Fig.~\ref{figs2}(a) and \ref{figs2}(b)]. Below $T_f$, the specific heat of $\alpha$-CrOOD also shows a power-law temperature dependence (see main text), which is insensitive to the applied magnetic field up to 4 T [see Fig.~\ref{figs2}(b)]. The freezing temperatures of both $\alpha$-CrOOH and $\alpha$-CrOOD detected by specific heat are consistent with those reported previously~\cite{Matsuo2006Isotope,maekawa2007deuteration-induced}. No obvious magnetic hysteresis is observed down to 1.9 K [see Fig.~\ref{figs2}(c)], the weak kink observed at low magnetic fields of $<$ 1 T in the whole temperature range is caused by the tiny magnetic impurity of $\leq$ 0.2\% from the stainless steel liner.

\begin{table}[t]
\caption{Rietveld refinements for $\alpha$-CrOOH with the delafossites structure (space group: $R\bar{3}m$). Here, Oc represents the occupancy fraction, Cr-Cr$_{intra}$ and Cr-Cr$_{inter}$ are the intralayer and interlayer distances of Cr-Cr, respectively.}\label{table2}
\begin{center}
\begin{tabular}{ l | l | l | l }
    \hline
    \hline
  & 300 K (x-ray) & 50 K (neutron) & 8 K (neutron) \\ \hline
$a$ ({\AA}) & 2.9814(2) & 2.9758(3) & 2.9739(3) \\
$c$ ({\AA}) & 13.4122(8) & 13.409(3) & 13.413(3) \\
$z$(O) & 0.40803(7) & 0.4060(2) & 0.4060(2) \\
$z$(H) (Oc: 0.5) & 0.506 (fixed) & 0.5001(3) & 0.5001(3) \\ \hline
$R_p$ (\%) & 4.22 & 2.65 & 2.73 \\ \hline
Cr-O ({\AA}) & 1.991 & 1.975 & 1.974 \\
$\angle$CrOCr ($^o$) & 96.92 & 97.75 & 97.74 \\
Cr-Cr$_{intra}$ ({\AA}) & 2.9814 & 2.9758 & 2.9739 \\
Cr-Cr$_{inter}$ ({\AA}) & 4.791 & 4.789 & 4.789 \\
    \hline
    \hline
\end{tabular}
\end{center}
\end{table}

\begin{figure}[t]
\begin{center}
\includegraphics[width=8.7cm,angle=0]{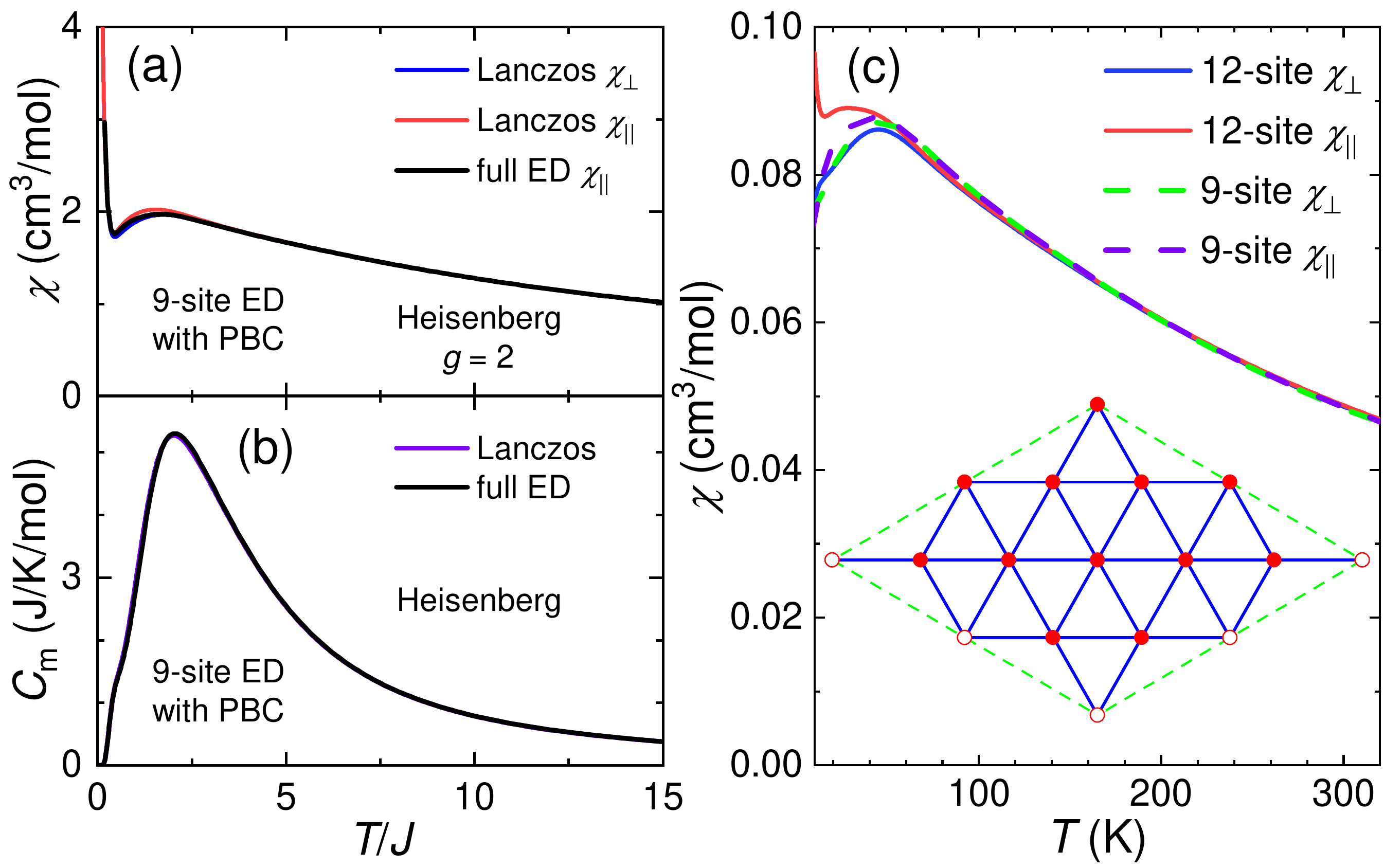}
\caption{(Color online)
Temperature dependence of (a) susceptibility and (b) specific heat calculated by Lanczos diagonalization and full ED techniques, based on the same $S$ = 3/2 9-site cluster with PBC and with the NN AF Heisenberg coupling of $J$. (c) Temperature dependence of susceptibility calculated by applying the 12-(solid) and 9-site (dashed) Lanczos diagonalization with the same set of easy-axis parameters, $J_1$ = 23 K, $D$ = -1.1 K, and $J_2$ = -2.3 K. The inset shows the 12-site cluster (solid) with PBC used in the FLD calculations.}
\label{figs3}
\end{center}
\end{figure}

The lattice specific heat (unit: JK$^{-1}$mol$^{-1}$) above $\sim$ 60 K is obtained by subtracting the calculated magnetic contribution ($C_m^{theor.}$) from the total measured specific heat for $\alpha$-CrOOH [Fig.~\ref{figs2}(d)]. Each CrO$_2$H has 3 acoustic and 9 optical modes, and thus we use the similar Debye-Einstein function as reported in Ref.~\cite{Matsuo2006Isotope} approximately,
\begin{multline}
C_{la}=\frac{9RT^3}{\Theta_D^3}\int_0^{\frac{\Theta_D}{T}}\frac{\xi^4e^{\xi}}{(e^{\xi}-1)^2}d\xi\\
+\frac{3R\Theta_{E1}^2}{T^2}\frac{e^{\frac{\Theta_{E1}}{T}}}{(e^{\frac{\Theta_{E1}}{T}}-1)^2}+\frac{6R\Theta_{E2}^2}{T^2}\frac{e^{\frac{\Theta_{E2}}{T}}}{(e^{\frac{\Theta_{E2}}{T}}-1)^2},
\label{eqs1}
\end{multline}
where $\Theta_D$, $\Theta_{E1}$, and $\Theta_{E2}$ are three fitting parameters (see below). By fitting to the lattice specific heat, we obtain $\Theta_D$ = 397(1) K, $\Theta_{E1}$ = 1507(50) K, and $\Theta_{E2}$ = 690(4) K, with the least-$R_p$ = 0.18 [please see Eq.~(\ref{Eq2}) for the definition of $R_p$]. With all three precisely fitted parameters at hand, we further calculate (extrapolate) the Debye-Einstein lattice specific heat using Eq.~(\ref{eqs1}) in the entire temperature range below $\sim$ 200 K [see Fig.~\ref{figs2}(d)].

\section{Quantum many-body simulation.}
\label{a2}

To check the precision of our FLD calculations for the $S$ = 3/2 TAF model, we first performed the full exact diagonalization (ED) calculation on the $N_0$ = 3$\times$3 = 9 -site cluster with PBC~\cite{PhysRevX.10.011007} in the ideal Heisenberg case ($J_1$ = $J$, $D$ = 0, and $J_2$ = 0). The dimension of the Hilbert space is given by $N_{st}$ = 4$^{N_0}$ = 262144, and the Hamiltonian matrix is too large for us to diagonalize directly. As $M^z$ $\equiv$ $\Sigma_jS_j^z$ (eigenvalue: $m^z$) and the six-fold rotational symmetry operation ($C_6^z$) commute with the THAF Hamiltonian, we can divide the large Hamiltonian matrix into smaller ones with the maximum size of 5086$\times$5086, through similarity transformations. After one-by-one diagonalizing these small Hamiltonian matrices, we obtain all of the eigenenergies, $\{ E_{m^z,j}\}$, with $m^z$ = -$N_0S$, -$N_0S$+1, ..., $N_0S$, at 0 T. Therefore, the partition function in a magnetic field of $\mu_0H^z$ applied along the $c$ axis is given by,
\begin{equation}
Z(T,\mu_0H^z)=\sum_{m^z=-N_0S}^{N_0S}\sum_je^{-\frac{E_{m^z,j}-\mu_0H^z\mu_Bgm^z}{k_BT}}.
\label{eqs2}
\end{equation}
We further calculate the spin susceptibility [black line in Fig.~\ref{figs3}(a)] and heat capacity [black line in Fig.~\ref{figs3}(b)] by,
\begin{equation}
\chi_{\parallel}=\frac{RT}{\mu_0N_0}\frac{\partial^2\ln Z(T,\mu_0H^z)}{(\partial H^z)^2}\mid_{H^z=0},
\label{eqs3}
\end{equation}
\begin{equation}
C_{m}=\frac{R}{N_0T^2}\frac{\partial^2\ln Z(T,\mu_0H^z=0)}{(\partial \frac{1}{T})^2}.
\label{eqs4}
\end{equation}

The full ED method requires extremely large CPU time, $\propto$ $N_{st}^3$, and memory, $\propto$ $N_{st}^2$, when the cluster size increases to $N_0$ = 12. Therefore, we choose to apply the FLD method to calculate the physical quantities using the Lanczos steps up to $M_L$ = 40 with the error of $O$(1/$T^{M_L+1}$) , as reported in Ref.~\cite{PhysRevB.49.5065}. First of all, we calculate the spin susceptibility and  heat capacity of the ideal $S$ = 3/2 THAF model using FLD method on the same 9-site cluster as ED, that are almost overlapped by the full ED results [see Figs.~\ref{figs3}(a) and \ref{figs3}(b)]. Then, we perform the FLD calculations on the 12-site cluster with PBC [see inset of Fig.~\ref{figs3}(c)] using the same $M_L$ = 40, and no significant size effect is observed above $\sim$ 60 K [see Fig.~\ref{figs3}(c)]. At each temperature, we first calculate the static magnetic moments at each site, $\langle$\textbf{S}$_j$$\rangle$, and then obtain the static structure factors [see Fig.~\ref{fig3}(b)] as,
\begin{equation}
|F_m|^2(\mathbf{Q})\propto\frac{1}{N_0^2}\sum_j\langle\mathbf{S}_j\rangle e^{i\mathbf{Q}\cdot\mathbf{r}_j}\cdot\sum_{j'}\langle\mathbf{S}_{j'}\rangle e^{-i\mathbf{Q}\cdot\mathbf{r}_{j'}},
\label{eqs4}
\end{equation}
where \textbf{r}$_j$ is the position vector of the $j$th site on the triangular lattice.

\bibliography{Cr_1}

\end{document}